# A Survey of Virtual Reality in Japan

Benjamin Watson
Graphics, Visualization & Usability Center
College of Computing
Georgia Institute of Technology
Atlanta, Georgia 30332
watsonb@cc.gatech.edu

## Abstract

The NSF Summer Institute in Japan program sends about 60 graduate students of all disciplines to Japan each summer. For two months, students participate in research at host labs, visit conferences and other labs of interest, and receive Japanese language and cultural instruction. Full financial support is provided by the American and Japanese governments. During the summer of 1993 the author participated in this program and took the opportunity to visit the Japanese virtual reality research community. He attended two virtual reality conferences and toured more than a dozen labs. After the program, he made short visits to VR and graphics labs in PR China and South Korea. This paper gives a detailed account of these experiences.

## 1. Introduction

Why is it that so many US researchers know so little about work in Asia? This author had always found this question a puzzling little mystery. It appeared that, especially in Japan, there was quality work to be found. Technologically and economically, there was no doubt that Japan was at least our equal in many areas. Was the research perhaps not as good as it seemed? Were we the victims of simple Western complacency? Or were we afraid of what we might find? It seemed that the answers to these questions would have to come firsthand.

The US National Science Foundation's (NSF's) Summer Institute in Japan was a chance to find these answers. This program provided the opportunity for an extended period of study and research in Japan, and shorter visits to the People's Republic of China and South Korea. The stay in Japan included two virtual environments conferences and more than a dozen visits to virtual environments or computer graphics labs. Most of the time in China and South Korea was spent attending two graphics conferences and visiting several labs doing computer graphics work.

This report describes the Summer Institute in Japan, and explains how other researchers may take advantage of it and similar programs. It then summarizes the visits made during the program, with a particular focus on those visits dealing with teleoperation and virtual environments. Because there is only a limited amount of virtual environments research being done in South Korea and China, visits in these countries are only briefly discussed. For a more comprehensive account of these travels, see (Watson, 1994).

Several others have reported on the state of VR research in Japan. Dr. Stephen Ellis of NASA described one of his many visits to Japan in (Ellis, 1992). Dr. David Kahaner of the US Office of Naval Research (ONR) has been in Japan since 1989 and publishes reports on his visits to labs and conferences around Asia electronically. His past reports are available via ftp. For a fairly recent index of them, see (Kahaner, 1993a). He recently described several Japanese VR projects in (Kahaner, 1994).

Reports on the general academic and scientific environment in Japan can be found in (Maddox & Swinbanks, 1992) and (Notkin & Schlichting, 1993). A similar report on South Korea can be found in (Swinbanks, 1993). For impressions of Japanese culture and society, please see (Mura, 1992) and (Christopher, 1984).

## 2. The NSF Summer Institute In Japan

Several years ago, around the time of President Bush's ill-fated trip to Japan with Lee Iacocca and other leaders of the American automobile industry, the US and Japan signed an agreement for scientific and technological cooperation (STA, 1988). As part of this agreement, a Task Force on Access was created to report on the degree of access researchers had to the American and Japanese research environments. In 1989, the Task Force compared the number of Japanese researchers visiting the US to the number of American researchers visiting Japan, and found ratios of 5.3 to 1 for short term visits, and 2.4 to 1 for long term visits (NSF, 1993a). This imbalance existed despite a long tradition of scientific cooperation between Japan and the United States: in 1961, President Kennedy and Prime Minister Ikeda had established a committee for scientific cooperation, the first of its kind in the world. Clearly the situation called for new initiatives.

The NSF Summer Institute in Japan program is just one of the ways in which the Japanese and US governments decided to address the problem. The program sends dozens of graduate students from US universities, working in a wide variety of fields, to Japanese laboratories for two summer months. There the students receive intensive Japanese language training, work and research in their host laboratories, attend various scientific and cultural lectures, and are encouraged to make research visits and attend conferences. The sponsors of the program hope that program participants will familiarize themselves with the Japanese research environment and form the sort of personal and professional relationships that lead to research collaboration. To this end, financial support for dissertation enhancement, post-doctoral research, and faculty visits are available from both the US and Japanese governments (STA, 1993; NSF, 1993b). Much of this support includes allowances for dependents.

The program began in 1990 with 25 graduate students visiting national laboratories at Tsukuba Science City, about 60 km northeast of Tokyo. In 1993 it had expanded to include 60 students, many of which were located in Tokyo proper. In addition, several corporate laboratories hosted students, the US National Institute of Health (NIH) supported several students in the biomedical field, and a trial program was started in Okazaki.

While the Summer Institute program is always filled to capacity, it is still relatively unknown. In 1993, applicants had roughly a 50 percent chance of being accepted. Refer to the appendix for information on applying to the Institute.

## 3. Conferences in Japan

This author was able to attend two conferences in Japan. Generally, Japanese conferences are slightly more formal than conferences in the United States and Canada. Business cards are a must. Meeting someone without one, especially when presented with one, can be very embarrassing.

### 3.1. IVR '93

The Industrial Virtual Reality conference (IVR, 1993) was held near Tokyo on June 23 and 24, 1993. As the name implies, the main emphasis of this virtual reality conference was industry and its products and applications. The conference consisted of talks, seminars, and an exhibition. About one-third of the conference talks and seminars were conducted in English.

The conference was divided into two sections: industrial applications and basic technologies. Several invited talks were given as well. This section briefly discusses the English portions of the conference, as well as a few of the more interesting Japanese talks. Hopefully, these latter talks will some day be available in English translation.

Professor Susumu Tachi gave the first invited talk, a general overview of virtual reality and teleoperation. Dr. Michitaka Hirose followed with a survey of previous and current virtual reality technology. Robert Mann of MIT was the third invited speaker, and reviewed his pioneering work in virtual environments (VEs) and the use of VE technology for aiding the disabled. The final invited talk was given by Dr. Tom Furness of the Human Interface Technology at the University of Washington. Dr. Furness did not include the text for his talk in the conference proceedings, only a two-page summary of his career -- which is interesting enough in its own right.

Most of the talks in the industrial applications section were given by representatives of Japanese corporations in Japanese. Two of the talks, however, were given in English. Satoru Nio of Yaskawa Electric Corporation described a semiautonomous robot with remote teleoperative interface under development at his company. Won S. Kim of JPL, a US research organization, discussed research in teleoperative controls going on at his lab. JPL is particularly interested in predictive interfaces for time delay environments. Interesting talks in Japanese in this section included Jiunji Nomura of Matsushita's presentation on VR in housing and electronics, Makoto Ishimitsu of NHK's discussion of VR and the future of broadcasting, and Mikio Okada's of Tokyo Power's overview of his company's work with VR and software visualization.

The first English talk in the basic technologies section was given by James Krieg of Polhemus, who discussed Polhemus' new FASTRAK technology. Charles Grimsdale of Division, Ltd. then described the VE programming environment dVS, and VCToolkit, a virtual reality toolkit. He was followed shortly by Pat Gelband of Sense8 describing the World ToolKit software. Later, Mark T. Bolas of Fakespace described the Boom2C. The final English talk in this section was given by Scott Foster of Crystal River Engineering. Mr. Foster discussed the basic challenges of generating 3-D sound. Interesting Japanese presentations included Hironao Tekeda of Sega's discussion of computer games and Koichi Otomi of Toshiba's overview current stereoscopic display technology.

IVR's technology exhibition was a disappointment. The exhibition space was clearly too small, resulting in a pedestrian traffic jam that raised both blood pressure and body temperatures.

There were only a few mildly interesting demonstrations, including the Japanese arm of Silicon Graphics with a Fakespace Boom, and Nissho Electronics, which was marketing VPL's VE hardware.

## 3.2. ICAT '93

The third International Conference on Artificial Reality and Tele-existence took place from July 6 to July 7 close to central Tokyo. This conference, though it included many of the same names as IVR '93, was much more academic in emphasis. Each day of the conference was divided into two sessions; each session began with an invited presentation, followed by the presentation of several submitted papers (ICAT, 1993a, 1993b).

The conference also included a student competition. Student groups from different universities were given access to the same equipment and equally sized budgets, and asked to create the best sorts of virtual interfaces and environments they could dream up. The students here created highly imaginative and sometimes frightening displays, some of which included electric shocks and chairs that moved at inconvenient moments. Despite the Rube Goldberg appearance of all the entries, demand for these virtual experiences was quite high, and it was no simple matter to gain a seat in one of the contraptions. All of the entries involved a lot of seat-of-the-pants engineering, the kind of work the typical American computer science student might find difficult. Japanese research in VR seems to come largely out of engineering departments, where hardware is a primary concern. In the United States, most VR research seems to be done by computer science departments, which are interested primarily in software and algorithms.

Short summaries of the conference's presentations follow:

### Dr. Carl Loeffler, CMU Center for Creative Inquiry

Dr. Loeffler is an artist interested in shared, interactive experiences and authorship, particularly over large distances. He is perhaps best known for his work with the Whole Earth bulletin board and his newer virtual museum project. He demonstrated a 486 PC-based virtual environment which allows two participants, connected only by a 9600 baud modem, to interact with each other in real time. In this case, one participant was at the conference in Tokyo, and the other at CMU. The environment, funded by Ford, allowed cooperative design of a Fiesta automobile. Another of Dr. Loeffler's current projects involves the design of a virtual city, which will be used in a forthcoming VR feature directed by George Romero, and which then will be developed into a full-
blown city simulation.

## Prof. Susumu Tachi, RCAST, University of Tokyo

Professor Tachi presented his impressive telepresence system. The system includes a homomorphic robot, with an arm and two video cameras that have all the degrees of freedom that the matching human body parts do, as well as a matching teleoperative interface, which includes a mechanically tracked head-mounted display and an exoskeleton for the operator's arm that allows a very intuitive control of the robot arm. The system is quite effective -- in one demonstration, the teleoperated robot delicately rearranged eggs.

Prof. Tachi presented an experiment with his system that measured teleoperative performance with different types of remote interfaces, including stereoscopic and monoscopic HMD displays, robot-mounted and traditional video displays, as well as direct observation. Users in the teleoperative environments were given the task of following a randomly moving marker with the robot arm. The marker had no side-to-side motion relative to the robot and was fairly close to it (about a meter away). Prof. Tachi's results indicated that next to direct observation, a stereoscopic HMD interface allowed the best performance.

## Haruo Takemura & Fumio Kishino, ATR Communications Systems Research Labs

Mr. Kishino presented the work of the department he heads at ATR. The main focus of the VE work he discussed was the use of VEs as a communication medium. One project allows two participants to communicate in a VE in real time. Participants can handle virtual objects of discussion and of course hear each other talk. Each remote site requires a large screen monitor, stereo glasses, and a dataglove. The user stands in front of the screen and sees the remote participant interacting with him in a VE. A video image of each participant's face is texture mapped onto the corresponding virtual face. Facial expression changes are based on the current facial expression of the user, which is tracked by placing small plastic dots on the participant's face and following them with a video camera. Only a limited number of virtual expressions were available, and the mapping from them to actual expressions was somewhat abrupt and crude. Low-resolution eye tracking was used to animate eye-motion and blinking.

Mr. Kishino quickly presented other work as well. In one ongoing project, hand shape is recognized using non-intrusive video technology. No gloves are required. Cameras are mounted over a desk allow the shape recognition in real time. No gestures were recognized, that is, no meaning had been given to any hand shape. Another project attempted to perform gaze tracking. Corneal reflection and marked glasses allowed the determination of eye and head orientation. Higher levels of graphical detail were provided on the screen at the center of fixation.

### Michael Zyda, Department of Computer Science, Naval Postgraduate School

Dr. Zyda presented his work on NPSNET, a large-scale land battle VE simulation compatible with DARPA's SIMNET. At the time of the conference, NPSNET could handle about 500 players. Ultimately, Dr. Zyda hopes NPSNET will be able to accommodate 10,000 to 300,000 players and 30,000 ground vehicles or 15,000 air vehicles on local area networks. NPSNET uses the DIS 2.0.3 VE networking standard.

Dr. Zyda had incorporated several hypermedia elements into NPSNET. As players move through the simulation, they encounter multimedia anchors, which can display graphics, audio, video, and text. These anchors can be triggered by proximity or explicitly by the user. Players can use the simulation itself to leave these anchors. Future uses of hypermedia in NPSNET include hypernavigation, to allow instant relocation within the simulation, and temporal anchors, which would only be visible during certain set time periods.

### Ryugo Kijima & Michitaka Hirose, Faculty of Engineering, University of Tokyo

Mr. Kijimia presented a psychological application of VE technology. The sand play (sandspiel) is a technique used in the treatment of autism. Patients play with toys and shape the sand, and psychologists form a diagnosis from the result. Mr. Kijima's system used a large screen stereoscopic display and a wand input device. The wand was represented in the VE by a similarly sized wand, and not an infinite half ray. Experiments with 40 patients showed that the system was useful in treatment. Mr. Kijimia indicated that one advantage of this treatment over traditional sandspiel therapy is that modern children prefer computer to true sandbox play.

### Gen Suzuki, Shouhei Sugawara & Machio Moriuchi, NTT Human Interface Labs

Mr. Suzuki presented a visual communication environment using VE technology. Users in this VE communication space are represented by upright, flat panels. The panels can move around like game pieces inside a virtual space (e.g., a lobby, a shopping mall). Video stills of the users are mapped onto each panel. The system was implemented on PCs networked with Ethernet. Since Ethernet does not have the bandwidth to texture map live video, users have the option of switching to normal videophone mode for animated video interaction. In the future, Mr. Suzuki hopes to port the environment to an ISDN network; its bandwidth should allow real time video texture mapping directly onto the users' panels.

### Kazuo Itoh, Asahi Electronics Co., Ltd.

In his talk, Mr. Itoh reviewed various VE platforms and devices. Among the many different types of hardware he discussed were a new HMD weighing half a kilogram with .7 inch LCD screens with 100K pixels. This he compared to the Virtual Research Flight Helmet (weighing 1.5 kg), the new Virtual Research Eyegen 3 (weighing .75 kg and with 1 inch CRTs with 370K pixels), the Sony Visortron, and an Olympus LCD projector (stereo, 648K pixels).

### Hiroshi Harashima, University of Tokyo

Dr. Harashima dreams of a new field of science, called "human communications engineering", which would refocus the attention of communications scientists from information transmission to the actual terminals: human beings. In line with this principle, he presented work on facial animation and face generation. His animation work began, as does so much facial animation research, with the Facial Expression Coding System (FACS) of Ekman and Friesen. Mr. Harashima has implemented a graphical version of FACS that allows the generation of facial expressions on a geometric facial model. It was not clear how fast this system could perform, or how accurately the system modelled the facial musculature. Dr. Harashima did admit that it was difficult to generate realistic wrinkles with his system. In order to allow structural as well as expressional facial variety, Dr. Harashima created a facial database, as well as a method of averaging the faces in the database to produce new faces. To close, he presented the average faces of 13 bank clerks, 11 professional wrestlers, and 10 political party faction leaders. For more details, see (Harashima & Kishino, 1991).

### Dr. Stephen Ellis, NASA Ames Research Center

Dr. Ellis addressed a great many issues. First he pointed out that VR is nothing new, that in fact, it is simply the extension of decades of flight simulation research, pioneered by Ed Link and Ivan Sutherland. Today's HMDs and other devices are just cheaper, slower versions of those developed years ago for flight simulators.  This poorer quality leads to what he views as many of today's research challenges: slow update rates, poor resolution, excessive weight, and poorly generated stereo images.

Dr. Ellis emphasized that virtual environments research is essentially human computer interface research, and that technical issues in the field must address the human issues.  As an example, he mentioned his recent research that showed that stereo computations for roll (side to side tilting of the head) are unnecessary, except at the extremes of head rotation. Simply allowing the VE user to turn while standing or providing a swiveling chair could negate even this effect.

Dr. Ellis also posed another important research question: how can simulation fidelity (what many call "immersion") be objectively measured?  He described his own research in this area, which used subjects' estimation of a gravity normal as such a measure.  This research found that incorporating textures into the VE improved fidelity.  He then mentioned Professor Howard of Canada's motion simulation research and Beth Wenzel's auditory displays research with as possible approaches to improving simulation fidelity.

**Hiroo Iwata & Hiroaki Yano, Institute of Engineering Mechanics, University of Tsukuba**

Dr. Iwata presented a project exploring the use of artificial life in VEs.  With his application, users can control the autonomous growth of a virtual tree.  The speed of the growth is controlled with VCR-like buttons.  Certain other widgets allowed the user to define congenital characteristics (e.g. branching frequency).  Finally, through the use of a custom force-feedback display, users can prune, graft, and replant branches of the tree.  Dr. Iwata and Mr. Yano plan to explore the more general use of artificial life and autonomy for construction of virtual worlds. They hope to allow the user to direct armies of small, simple-minded artificial life forms.

**Juli Yamashita & Yukio Fukui, National Institute of Bioscience and Human Technology, AIST, MITI**

Ms. Yamashita presented a new technique for deforming free form surfaces.  She claimed that the technique was much more intuitive and interactive than deformation through control points, physical modelling, or surface geometries.  The implementation of her technique is very simple: a weighted version of the vector of the user's latest interaction is added to the surface's control points.  A width parameter controls the extent of the deformation's effect.  Ms. Yamashita presented both 2D and 3D implementations of her technique.  The 3D implementation made use of a force feedback display under development at her lab.  Ms. Yamashita hopes to add a suite of deforming "tools" to her implementation, in effect allowing 3D sculpture.

## Masahiro Ishii & Makoto Sato, Precision and Intelligence Laboratory, Tokyo Institute of Technology

Mr. Ishii presented a force-feedback device.  The device took the form of a frame, in the shape of a cube, a little less than one meter across on each side.  Within the cube were two finger-sized rings, each of which were attached to the frame by four thin wires.  Users could place index finger and thumb into these rings, and the tension on the wires could be varied to give force feedback.  An experimental comparison virtual and real wooden block manipulation showed that the device displayed force realistically. Mr. Ishii pointed out that care must be taken not to tangle the wires when using the display.  See video (TCAR&T, 1993) for a demonstration of the system.

## Martin Buss & Hideki Hashimoto, Institute of Industrial Science, University of Tokyo

Mr. Buss presented a force feedback display and system under design as part of an effort to provide intelligent, cooperative assistance to machine users.  Their display will "learn" manipulation techniques from the user, remember them, and then either apply them directly in fully automated mode, or use them to assist a user during a different manipulation.

## Discussion Panel: Susumu Tachi, Michitaka Hirose, Michael Zyda, Stephen Ellis

This very interesting discussion began with Dr. Zyda answering the question: what direction is the VR research in the States going to take? Dr. Zyda pointed out that Congress had asked the National Academy of Science to produce a report addressing exactly this issue.  Mr. Durlach of Presence and Dr. Zyda himself are involved in this effort.  The report is scheduled for publication as a book in 1994.

Addressing Dr. Ellis' earlier presentation, Dr. Zyda asserted that what is new about VR is the fact that so many old technologies are being integrated.  The glue for this integration, software, is a new development. Also, physics was not previously so completely integrated into software.

The panelists agreed that two great needs in VR are a driving problem, what Dr. Ellis called " a VisiCalc for VR," as well as a set of VR standards.  Dr. Tachi expressed the belief that "natural selection" in the industry will provide these standards. Dr. Zyda also stressed the need for real-time architectures, with large, shared memory spaces. He believes that new techniques for polygon flow minimization are required.

## 4. Visits in Japan

Geography makes visiting the Japanese research community simpler than it might otherwise be: most institutions of significance are located either in the Tokyo or Kyoto/Osaka areas. The government is trying to change this, but this is still a fact of Japanese life. With one exception (ATR), every lab described in this section is located in the Tokyo area.

### 4.1. Dr. Hirose at the University of Tokyo

Dr. Hirose's lab is very productive, with many researchers involved in a wide range of virtual reality research. This section describes just a few of the lab's many projects.

The product of Koichi Hirota's research is a force display (Hirota, 1993). A small ring magnet is placed on the user's finger, which in turn is placed inside a small metal box. A larger, surrounding device moves the box as the finger moves. To display force, the finger is allowed to come into contact with the box. The finger could only be moved slowly, or else the box would lag behind. A stereoscopic graphic display is synchronized with the force feedback device.

The wind display is a small, lightweight plastic box. At the center of the box is a handle. Four small fans in the box surround the handle. As the user grasps the box, and views a three-dimensional vector field, the strength and direction of the vectors close to the virtual cursor are felt as air currents on the hand. The end result is very effective.

Kensuke Yokoyama is working on a research project called the "virtual dome" (ICAT 1992; Kahaner 1992a; Hirose 1991a, 1992a). The basic idea is simple: it is very difficult to present realistic looking remote or virtual environments, and at the same time allow real time response. Mr. Yokoyama solves the problem by texture mapping video images onto a polygonal dome surrounding the user. This approach allows the user to use realistic head motion to view the environment. The video texture maps were not themselves updated in real time. A later implementation uses computer vision stereo range-finding techniques to automatically produce three dimensional models onto which the video is mapped. The end result is a very realistic display of a three dimensional still image.

Dr. Hirose's lab does work in three-dimensional sound, augmented reality (Hirose 1991b), fine work in software visualization (Yamashita 1991; Hirose 1990a, 1991b, 1991c), and human interaction in virtual environments (Hirose 1990b, 1992a, 1992b). It is also exploring the virtual sandbox application described above (ICAT 1993a). For an account of an earlier visit to Dr. Hirose's lab, see (Kahaner 1991). See also video (ICAT 1993b).

## 4.2. Mr. Suzuki & Mr. Sugawara at NTT

Mr. Suzuki's Visual Communication Environment Group at NTT's Human Interface Laboratories is attempting to turn VR technology into communication technology. Mr. Sugawara demonstrated the virtual communication environment Mr. Suzuki presented at ICAT (1993a). This demonstration made it clear that at NTSC resolution, it was quite to identify other users in the environment who were not fairly close.

A newer system allowed remote collaboration while handling an actual object (Suzuki, 1993). Each end of the system is equipped with a normally mounted monitor. In front of those monitors, underneath plexiglass tables, are additional monitors. Over the tables reach a robot arms, holding small paddles. Each player stands in front of a table. Player 1 grasps a robot arm, a master to the slave robot arm at the other table. Player 2 holds a real paddle. A small ball is dropped in front of player 2 and the game begins. On the normally mounted monitors, each player sees images of himself and his opponent. Player 2 sees the actual ball, his paddle, and the slave robot arm's paddle. Player 1 sees the same scene displayed by the monitor below the table before him. By moving his master robot arm, he and player 2 can play a game with an actual ball, even though they may be widely separated. Mr. Suzuki stated that several arcade companies had expressed interest in this system. He also envisions applications for computer shopping and education. Work at Mr. Suzuki's lab on a collaborative workspace was presented at (ICAT 1992; Kahaner 1992a).

## 4.3. Mr. Kishino at ATR

Advanced Telecommunications Research (ATR, 1993), outside of Kyoto, is an eight-year old research institution established by the government and NTT. Now it serves as something of a research consortium. About 200 researchers from a host of different companies, including NTT, the Japanese international telephone companies, NHK and others come as visiting researchers for a number of years.  After their tenure, they return to their original companies.  About 10 percent of ATR's researchers are foreigners, who come primarily from academic settings.

Dr. Haruo Takemura presented the virtual space teleconferencing environment (Takemura & Kishino, 1992) that Dr. Kishino, director of the Artificial Intelligence Department at the Communications Research Laboratories (ATR, 1992), had presented at ICAT '93.  The explained that the expression of the virtual face is texture mapped onto a model created off-line with a 3D scanner.  To allow virtual simulation of eye movement, the user must wear a helmet mounted with two small video cameras, which point at the eyes.  Each of the virtual eyes could only be in ten different positions.

ATR's real-time hand-shape recognition system tracks the orientation of the wrist and the positions of the fingertips.  The system, which truly does function in real-time, works over highly cluttered desktops.  One problem still being solved was occlusion of a hand by another hand or simply by improper perspective.  The work was being extended to allow simultaneous recognition of the shape of two hands.

For a summary of the virtual environments work at Mr. Kishino's department, see (Harashima & Kishino, 1991).  For an account of previous visits to ATR, see Dr. Kahaner's reports (1992b, 1994).

### 4.4. Mr. Kotoku at MEL

The Mechanical Engineering Laboratory (MEL) is one of the several national laboratories located in Tsukuba.  Like most such institutions, MEL contains several departments and laboratories.  Tetsuo Kotoku works in the Cybernetics Division of the Robotics Department, and demonstrated his force display (Kotoku et. al., 1992a, 1992b).  The display takes the shape of a manipulable mechanical arm with four degrees of freedom (two rotational axes were left out for simplicity).  The arm is connected to a PC-based transputer, which performs control and interfaces with a graphics workstation that displays a virtual block world.

The user could use the arm to move one virtual block among others.  The system was very effective.  There did not seem to be any delay in the feedback, and it was quite simple to differentiate between surface-surface contacts and surface-edge contacts.  Mr. Kotoku plans to increase the complexity of his system's virtual world. At the time of the demonstration, such worlds introduced feedback delay.

## 4.5. Prof. Tachi at the University of Tokyo

Professor Tachi, like Dr. Hirose, is affiliated with the University of Tokyo. However, Professor Tachi is not located on the university's main campus. His lab is at a western research campus, called the Research Center for Advanced Science and Technology (RCAST). Dr. Tachi is a pioneer in the field of telepresence in Japan. He chairs of the Technical Committee on Artificial Reality and Tele-Existence, a loose confederation of labs at several national institutes and universities that perform VR-related research (TCAR&T, 1993).

Dr. Tachi 's teleexistence system, which he had presented at ICAT '93, was very impressive. Several visitors, despite being first-time users, were able to use the system to stack small wooden blocks at the remote location. Dr. Tachi had also designed virtual equivalents of his robot and lab. Users could use the telerobotic interface to stack blocks in the virtual lab. Dr. Tachi envisions using this telepresence system for work in hazardous environments. He hopes to use his virtual system for environments that include significant time lag.

Dr. Tachi's students are involved in a wide range of projects. Takashi Oishi is working on a transparent HMD that will use gaze direction as an object reference. Satoru Emura is combining a Polhemus-like sensor with an inertial tracker to reduce positional lag. Michiko Okura researching auditory horopters, and Akihiko Hayashi is working on force feedback. For some details on Dr. Tachi's work, see (ICAT, 1992; Kahaner, 1992a; Tachi, 1984, 1991, 1993).

## 4.6. Dr. Iwata at the University of Tsukuba

Dr. Hiroo Iwata and his students at the University of Tsukuba demonstrated the system Dr. Iwata had presented at ICAT '93, as well as several other very interesting projects.

Haruo Noma's six degree of freedom force display (Iwata & Noma , 1993) took the shape of a joystick mounted on a complex of levers and gears. In one demonstration, users could use the display to move a 3D cursor against a cube-shaped object, and feel its edges and corners. In another application of the device, two spherical representations of molecules floated in space. As the cursor moved through one molecule, the strength of its bonds could be felt as resistance, pushing away from the molecules' center. As the cursor through the other molecule, its bonds could be as torque on the joystick. In a final demonstration, the Mr. Noma's device was used to shape a flexible surface. The display was a very effective way of indicating surface contact.

In Keigo Matsuda's haptic walkthrough system, users slipped a large flexible band around their waists, and put on two roller-skates. Since the flexible band was fixed to a large frame, the user could walk in place, while small trackers attached to the skates measured the length of his stride. Users turned by striding at an angle. In (ICAT, 1992; Kahaner, 1992a), an experiment is described that showed that this interface allowed better estimation of the virtual distance travelled than did simple virtual flying. All of the above projects are demonstrated in (TCAR&T, 1993), and some are discussed in (Kahaner, 1994).

### 4.7. Mr. Mitsuhashi of NHK

Mr. Mitsuhashi works at the Visual Science Research Division of NHK's Science and Technical Research Laboratories (NHK, 1992, 1993). His division works on new display technologies and the accompanying basic psychophysical research. One prototype is a large-screen, rear-projection, lenticular stereoscopic display. The prototype is very effective, once the optimum viewing position is found. To heighten the effect, the viewer is seated in a special surround-sound chair, embedded with speakers. NHK is betting that "no-glasses" stereo will be the future of binocular display.

Nobuyuki Hiruma is doing some very interesting research on the relation between the accommodative and vergence responses (1991). A long-standing worry of the VR community has been the conflict between stereo (vergence) input and screen distance (accommodation) input. Results of Mr. Hiruma's work indicate that as long as stereo input indicates distances within the viewer's current depth of focus, the accommodation response is coupled to the stereo input, and there is no conflict. This suggests a guideline for stereo viewing conditions, and by extension for stereo image generation.

Mr. Mitsuhashi's lab is also attempting to characterize random eye movement. This should ultimately have implications for gaze-tracking. NHK is heavily involved in research on auditory display, as well as a project called SSAV (super-surround audio-visual), which NHK hopes will be the next step after high definition TV.

### 4.8. Prof. Kawahata of Fujitsu

Professor Masahiro Kawahata, director of the Fujitsu Research Institute, showed me the Virtual Vision Sport, a new sort of HMD from a US startup company with which Dr. Kawahata is involved.  The Sport looks like a fairly large pair of glasses.  Over the right eye, however, is a small light that projects through an LCD screen onto the lower part of the right lens.  The picture is small and of fairly low resolution, but is full color and certainly sharp enough to serve its purpose.  The glasses are connected to a battery pack and broadcast television receiver, which can fit into most pockets.  The product was recently reviewed in Consumer Reports (1993).

Dr. Kawahata then introduced me to the Greenspace Project (Kawahata, 1993), an effort by Dr. Furness' Human Interface Technology Lab and Dr. Kawahata in Japan to create what might be called a virtual reality equivalent of the Internet (on a much smaller scale).  The Greenspace Project will connect the US and Japan to one, culture-spanning virtual environment. The project is envisioned as both a technological and cultural boot-strapping device. Dr. Kawahata indicated that the project is open to additional American or European partners.  One of the organizational umbrellas for this project is the Virtual Worlds Society, which at present includes members from the US and the Japan.  The mission of this society is to develop and apply new technology that will "link minds globally" and "unlock the power of human intelligence."

Dr. Kawahata is involved in a wide range of research, and as such holds several academic positions: he is a professor at Tokai University, an adjunct professor at the University of Southern California, and an affiliate professor at the University of Washington.  For an interview of Dr. Kawahata and a survey of the Japanese cultural VR scene, see (Grey, 1993).

### 4.9. Dr. Oka at RWC

Dr. Ryuichi Oka works with the Real World Computing Partnership (RWC).  The RWC (1992, 1993) is one of MITI's national projects, and is commonly viewed as the successor to the Fifth Generation Project.  The general goal of the project is to establish a human-like, flexible information processing system.  Research will concentrate in five areas: realizing human-like information processing; achieving a real-time, flexible, interactive understanding and learning capability (virtual reality would fit in here); creating massively parallel systems; developing large- scale neural systems; and establishing optical technologies for large-volume information transmission and processing.  The project, which started in July of 1992, will continue for ten years.  For the project's first five-years, the budget is $50 million.

Dr. Oka's department is located in the Tsukuba Mitsui building, the tallest building in Tsukuba. Several of its floors were donated to the RWC by one of the project's corporate partners. Dr. Oka's Theory and Novel Functions department is well equipped with, among other things, a Connection 64, an SGI Crimson and several Indigos, Sun and HP workstations, as well as two robots, a frame grabber, vision processor, and a large screen display.

Dr. Oka is primarily interested in motion-based recognition techniques (Oka & Takahashi, 1993). He and Susumu Seki demonstrated their global gesture recognition system. A user stands in front of a plain background. Through a video camera, Dr. Oka's system recognizes gestures such as clapping, head scratching, and waving. The system is very interactive; recognition is performed in real time with minimal delay. This technology seems perfectly suited for the virtual environments, and brings to mind Myron Krueger's full-body gesture recognition artificial reality interface.

## 4.10. Mr. Hashimoto & Mr. Yamamoto of Sony

The motto at Sony Corporate Research Laboratories in Tokyo is "research makes the difference" (Sony, 1993). Yoshitaka Hashimoto heads the Hashimoto Signal Processing Laboratory (Hashimoto, 1993). The research there is wide-ranging, and includes digital signal processing, high speed digital circuitry, computer graphics, and image recognition. Mr. Hashimoto demonstrated a system for real time animation (Kirkpatrick, 1991). The system's custom hardware can produce a very impressive real-time facial animation by loading a model of a Noh (ancient Japanese theater) mask, and saving a series of morphed facial expressions on the mask. When text is input to the system, the face "reads" by mapping certain key facial transformations to textual phonemes.

Mr. Masanobu Yamamoto is general manager of the Display Technology Research Department. He and Hiroshi Yoshimatsu demonstrated two HMD prototypes. One was similar to the Sony Visortron. It was extremely light, consisting of a headband and a small electronics unit that covered the eyes and contained two color LCD screens. Resolution was adequate, but not impressive. The display had a fairly narrow (about 60 degrees) combined field of view and was not true stereo; it was meant for use with normal monoscopic video signals. It was completely opaque, and included adjustments for both focus and interpupillary distance. No head tracking was performed. The prototype was essentially a head-mounted television, which was quite in line with Sony's expected applications: entertainment in airplanes and dentist's offices. The second prototype was functionally quite similar to the first HMD, with one major exception: it could be either opaque or transparent.  This product was obviously much farther from the market than the first.

Mr. Yamamoto was very anxious to hear of any studies on the effect of prolonged HMD usage.  His own preliminary studies had indicated that two hours of use of his HMDs did not lead to any ill effects.  However, his HMDs are quite light and do not incorporate motion tracking, and so might prove to be much less problematic than a tracked HMD configuration.

## 4.12. Mr. Akamatsu at the Inst. Bioscience & Human Technology

The Institute of Bioscience and Human Technology (IBHT) in Tsukuba is another MITI national laboratory.  It is involved in a wide and fascinating range of research related to the relationship between humans and technology.  Seeing it all would probably take several days.

Dr. Motoyuki Akamatsu began studying human fusion of different sensory modalities in the hope that the knowledge gained would be useful for robotic sensory fusion.  His studies concentrated on the fusion of the tactile and visual senses, and showed that the combination of these senses seemed superior to vision alone (Akamatsu, 1992).  His work then evolved into interface device design, as his colleague Sigeru Sato and he created a mouse with tactile and force feedback as a way of exploring his early results.  One of the mouse's original two buttons was replaced with a small, rounded pin that could poke out of the mouse casing to give tactile feedback.  A ferrous mouse pad allowed magnetic force feedback. Experiments showed that the device improved performance during use of normal 2D windowing computer interfaces (ICAT, 1992; Kahaner, 1992a).

The device is very simple.  It connects to any PC through a normal serial port.  It is powered by a standard AC converter.  This simplicity does not seem to harm its effectiveness.  Surprisingly, in the two years since its development, Dr. Akamatsu has had many commercial nibbles, but no bites.

Mr. Sato demonstrated another device under development.  The device, which looks like a glove, measures and records force intensity and distribution on the human hand (Sato, 1992).  It is envisioned for systematic study of human hand use.  The device can measure force at 15 different locations on each finger, 9 different places on the thumb, as well as several other locations on the palm.  Each of these force sensors rates force on a scale of one to seven, and outputs the result as color to a graphical representation of the glove.  Julie Yamashita demonstrated the work she had presented at ICAT '93. The deforming program seemed as intuitive as advertised, although the six degree of freedom force-feedback display (under development by Yukio Fukui with industry) still had some bugs.

Makoto Shimojo and IBHT are also involved in a MITI national project that should be of great interest to virtual environments researchers. The project is titled "Human Sensory Measurement Application Technology", and will attempt to realize simple and quantitative measurement and evaluation of the human senses (Shimojo, 1992). Part of this project will be the creation of simulated environment presentation technology. The project, which began in 1990, will continue until 1998, and has a $200 million budget. For demonstration and discussion of this and other work at IBHT, see (TCAR&T, 1992; Kahaner, 1994).

### 4.12. Mr. Bill Farmer at Sharp

Bill Farmer works at the Integrated Media Laboratories of Sharp Corporation. Mr. Farmer's lab had a very open and Western research style. Unlike most other labs, a simple eight hours of work each day seemed satisfactory. The lab is young and a newcomer to the VE field. It has three sections: one concentrating on optical engineering, one on virtual environments technology, and one concentrating on the human computer interface. Sharp, as a leader in LCD technology, has recently decided to jump on the HMD bandwagon. It has embarked on two major projects. The goal of one is to reduce the aliasing present in LCD displays. The other is to produce an 80 degree field of view HMD with minimal (or no) distortion.

Mr. Farmer, who works in the optical engineering section, is part of the design team for the new HMD's lenses. He is very interested in software techniques for eliminated optical distortion. If software could take care of distortion, Mr. Farmer and his fellow designers would be free to worry about other lens parameters. The virtual reality section is well-equipped with two SGI Skywriters, several X machines, a few custom-built booms, a VPL DataGlove and HMD, and a complete video editing suite.

### 5. Other centers of VR research in Japan

There are many labs and companies in Japan performing VR research not discussed in this report. NEC in Kanagawa near Tokyo is working with cooperative, networked virtual environments. They published some of their work in (ICAT, 1992), and Dr. Kahaner reviewed their research in (1992a) and (1994). Matsushita has long been famous for their work with a virtual kitchen. Their system allows customers to visualize and choose from various kitchen designs (Nomura et. al., 1992; Kahaner, 1994). According to Matsushita, 70 percent of the customers who use the system actually buy a kitchen, compared to only 35 percent of the customers who do not use the system. Finally, Fujitsu has been doing some interesting research with something called "virtual animals".

## 6. Prof. Wohn at KAIST in South Korea

Despite attending an excellent computer graphics conference (Pacific Graphics, 1993; Kahaner, 1993b) and visiting two technical universities in South Korea, I found only one lab in South Korea involved in virtual reality research. The lab is headed by Professor K. Wohn of the Korean Advanced Institute for Science and Technology (KAIST). KAIST is South Korea's most respected technical university. Many expatriate Korean researchers with decades of American experience have returned to South Korea and KAIST to establish laboratories, with ample financial support from private industry and the South Korean government.

Dr. Wohn has a small networked collection of PCs, which he plans to expand shortly with the addition of some Silicon Graphics equipment. SGI equipment is becoming increasingly common in South Korea. Work at his lab is going on in three areas: virtual reality, computer music, and computer vision. Dr. Wohn believes these areas are related. Specifically, he plans to attack the VR frame rate problem with techniques and psychophysical principles gleaned from computer vision. At the moment, however, his students are concentrating on the development of a PC-based, networked VR application environment.

## 7. Asian and American VR

As this report makes clear, the virtual reality community in Japan is quite large, and its researchers' investigations are compelling and of a quality comparable to work in the United States. Yet there are some fundamental differences between the Japanese and American efforts. First, virtual reality research in Japan seems to have an engineering emphasis, while in the States it has a computer science emphasis. Most of the virtual environments groups discussed in this report are heavily involved in the construction of one-of-a-kind devices. In the States, on the other hand, the research focuses more on software and the human-computer interface. Second, most virtual environments research at Japan's public research institutes focuses on telepresence and force feedback, while private institutes seem to concentrate on telecommunications and display technology.  American research shows more breadth, and includes telepresence and telecommunication as well as visualization, medicine, simulation, training and psychophysics.

What might explain these differences?  Science in Japan has traditionally emphasized engineering and applied research.  Though this observation is now by no means as valid as it once was, it provides some explanation for the Japanese VR community's engineering focus. Furthermore, while virtual reality is a young field, Japan's virtual environments research efforts are, on the whole, even younger than those in the US  Since robotics and electronics are already Japanese specialties, the related applications of force feedback, telepresence, telecommunications and display technology were natural starting points for Japanese VR researchers.  Indeed, some of the oldest VR labs in Japan emphasize telerobotic technology.  Newer Japanese projects and labs have begun to expand the scope of Japanese VR research.

Although this author's visits in South Korea were very short, they were nevertheless very informative and helped to put the Japanese VR research environment in perspective.  While Chinese interest in computer graphics is strong (CAD/Graphics, 1993), researchers there lack the technical infrastructure required for virtual reality work.  South Korean researchers are just beginning to take a serious look virtual reality, and should be producing significant results within a few years.  Despite this, Japan should remain the center for virtual environments research in Asia.  For a more complete report of this author's travels in China and South Korea, see (Watson, 1994).

## 8. Conclusion

Japan is doing important and meaningful virtual reality research, and researchers who make the effort to become familiar with the work going on there can only benefit.  There are obstacles to this process; in particular, because mastering written Japanese can take years, it is very difficult to keep abreast of scientific articles not written for international audiences.  Nevertheless, even secondary sources such as this report are useful, and often contain contact information that allows follow-up with the researchers themselves.  In my experience, Japanese researchers are quite willing to answer questions about their work.  Of course, especially in the field of virtual reality, nothing can replace an actual demonstration, and the NSF Summer Institute in Japan and similar post-graduate and visiting scholar programs are excellent opportunities to gain this sort of first-hand experience.

## 9. Acknowledgments

Dr. Kazuhiko Yamamoto, Dr. Shigeru Muraki, and the entire Image Understanding Section at the Electrotechnical Laboratory were my hosts in Japan and instrumental in arranging many of my visits.  Mr. Gentaro Hirota of IBM Japan and Dr. Tetsuo Kotoku of MEL not only hosted me at their own labs but also arranged visits with to Prof. Hirose's and Prof. Tachi's labs.  Ms. Janice Cassidy of NSF and Meg Nakamura of JISTEC provided me with the opportunity to visit Japan and invaluable guidance while there.  I am very grateful to them all for their help; my travels would not have been the success that they were without them.  A simulation fellowship from the Link Foundation provided financial support while this report was being written.

## 10. Contacts

*NSF Summer Institute in Japan:*

Japan Program
Division of International Programs
National Science Foundation
4201 Wilson Boulevard
Arlington, VA 22230
Tel: (703) 306-1701
Fax: (202) 653-5929
Email: NSFJinfo@nsf.gov

*Hirose's lab at University of Tokyo:*

Dr. Michitaka Hirose
Dept. Mechano-Informatics
Faculty of Engineering
University of Tokyo
3-1 7-chome Hongo Bunkyo-ku
Tokyo 113, Japan
Tel: (3) 3812-2111 ext. 6367
Fax: (3) 3818-0835
Email: hirose@ihl.t.u-tokyo.ac.jp

*Gen Suzuki at NTT:*

Gen Suzuki
Visual Communication Envt. Grp.
Visual Media Laboratory
NTT Human Interface Laboratory
Nippon Telephone and Telegraph Co.
1-2356 Take Yokosuka-shi
Kanagawa 238-03 Japan
Tel: (468) 59 2946
Fax: (468) 59 2829
Email: gen@nttvdt.ntt.jp

*Fumio Kishino at ATR:*

Fumio Kishino
Artificial Intelligence Dept.
Communication Systems Rsrch. Labs.
Advanced Telecommunications Rsrch Inst. (ATR)
2-2 Hikaridai
Seika-cho Soraku-gun
Kyoto 619-02 Japan
Tel: (7749) 5 1260
Fax: (7749) 5 1208
Email: kishino@atr-sw.atr.co.jp

*Tetsuo Kotoku at MEL:*

Tetsuo Kotoku
Cybernetics Division
Robotics Department
Mechanical Engineering Lab.
Ministry of Industrial Trade & Industry
Tsukuba Science City
Ibaraki, 305 Japan

*Dr. Susumu Tachi at University of Tokyo:*

Prof. Dr. Susumu Tachi
Rsrch. Ctr. for Advanced Sci. & Tech.
University of Tokyo
4-6-1 Komaba, Meguro-ku
Tokyo 153 Japan
Tel: (3) 3481 4467
Fax: (3) 3481 4469

Tel: (298) 58 7284
Fax: (298) 58 7201
Email: m4460@mel.go.jp

Email: tachi@tansei.cc.u-tokyo.ac.jp

*Hiroo Iwata of University of Tsukuba:*

Dr. Hiroo Iwata
Institute of Engineering Mechanics
University of Tsukuba
Tsukuba, Ibaraki, 305 Japan
Tel: (298) 53 5362
Fax: (298) 53 5207
Email: iwata@kz.tsukuba.ac.jp

*Tetsuo Mitsuhashi at NHK:*

Tetsuo Mitsuhashi
Visual Science Research Division
Science and Technical Resrch. Labs.
NHK (Japan Broadcasting Corporation)
1-10-11, Kinuta, Setagaya-ku
Tokyo 157 Japan
Tel: (3) 5494 2362
Fax: (3) 5494 2371

*Dr. Kawahata at Fujitsu:*

Prof. Dr. M. Kawahata
Fujitsu Research Institute
Fujitsu Limited
FACOM Bldg., 21-8, Nishi-Shimbashi
3-chome, Minato-ku
Tokyo 105, Japan
Tel: (3) 3437 3271
Fax: (3) 5472 5683

*Dr. Oka at RWC:*

Theory and Novel Functions Department
Real World Computing Partnership
Tsukuba Mitsu Bldg. 17F
1-6-1 Takezono
Tsukuba-shi, Ibaraki 305 Japan
Tel: (298) 53 1647
Fax: (298) 53 1640
Email: oka@trc.rwcp.or.jp

*Mr. Yamamoto at Sony:*

Masanobu Yamamoto
Display Technology Rsrch. Dept.
Corporate Research Labs.
Sony Corporation
6-7-35 Kitashinagawa
Shinagawa-ku, Tokyo 141 Japan
Tel: (3) 5448 2945
Fax: (3) 5448 7907

*Dr. Akamatsu at IBHT:*

Dr. Motoyuki Akamatsu
National Inst. Bioscience & Human Technology
1-1 Higashi, Tsukuba
Ibaraki 305 Japan
Tel: (298) 54 6772
Fax: (298) 54 6757
Email: akamatsu@nibh.go.jp

*Mr. Farmer at Sharp:*

Bill Farmer
Integrated Media Laboratories
Multimedia Systems R & D Center
Corporate R & D Group
Sharp Corporation
1-9-2 Nakase, Mihima-ku
Chiba 261 Japan
Tel: (43) 299 8700
Fax: (43) 299 8709
Email: farmer@iml.mkhar.sharp.co.jp

*Dr. Wohn at KAIST:*

Dr. Kwangyeon Wohn
Computer Science Dept.
Korea Advanced Inst. of Science & Tech.
37301 Kusong-dong, Yusong-ku
Taejon 305-701 S. Korea
Tel: (42) 869 3532
Fax: (42) 869 3510
Email: wohn@cs.kaist.ac.kr

**Appendix: Applying to the NSF Summer Institute in Japan**

While the NSF Summer Institute in Japan program provides assistance in locating a host laboratory, the best host/student matches come when the student is involved in the lab selection process. The NSF requests applicants to provide a list of appropriate host laboratories; and applicants should not hesitate to approach potential hosts themselves. A major criteria in finding a match is proximity to Tokyo and Tsukuba, since that is where students are housed. Graduate students eligible for the Summer Institute must be US citizens or permanent residents. No prior knowledge of Japanese is required. The sponsors do not recommend that spouses or dependents come to Japan with program participants (NSF, 1994).

The Institute is jointly administered by the NSF on the US side, and the STA (Science and Technology Agency, a division of MITI) on the Japanese side. Funding comes from NSF and NIH in the US. In Japan, the Center for Global Partnership (CGP) and Japan International Science and Technology Exchange Center (JISTEC) provide financial and logistical support. US funding covers students' travel costs to and from Japan, as well as an extra $2000 as compensation for income lost during the program's duration. Japanese funding covers room and board, costs of research travel, as well as several enjoyable cultural experiences. In my experience, the amount of money provided to each student is more than adequate.

Interested students may request an application form electronically or by post. For contact information, see the contact list earlier in this report.